\documentstyle[prl,aps]{revtex}

\begin{document} 
\input epsfig.sty      
%\draft

\title {
  A Numerical Study of  Ultrametricity
  in Finite Dimensional Spin Glasses}

\author{A. Cacciuto, E. Marinari}
\address{Dipartimento di Fisica and Infn, Universit\`a di Cagliari\\
Via Ospedale 72, 09100 Cagliari (Italy)}

\author{G. Parisi}
\address{Dipartimento di Fisica and Infn, 
Universit\`a di Roma {\em La Sapienza}\\
P. A. Moro 2, 00185 Roma (Italy)}

\date{\today}

\maketitle

%\widetext
 
\begin{abstract}
We use a constrained Monte Carlo technique to analyze ultrametric 
features of a $4$ dimensional Edwards-Anderson spin glass with 
quenched couplings $J=\pm 1$. We find that in the large volume limit 
an ultrametric structure emerges quite clearly in the overlap of 
typical equilibrium configurations.
\end{abstract}
 

\narrowtext

The hierarchical solution \cite{PARISI} of mean field spin glasses 
\cite{PARBOO} introduces a large number of new features. For 
$T<T_{SG}$, in the broken phase, there are many stable equilibrium 
states, not related by a simple symmetry group (like $Z_2$ in the 
normal Ising model). These states exhibit an ultrametric structure: 
their distance satisfies an inequality stronger than the triangular 
inequality, $d_{1,3} \le d_{1,2} + d_{2,3}$, the ultrametric 
inequality, stating that  $d_{1,3} \le \max(d_{1,2},d_{2,3})$. 
The existence of a phase transition even in non-zero magnetic field 
(de Almeida-Thouless line) and of a complex dynamics, with aging 
phenomena, are other crucial features of this picture.

The question of main interest is how many of such remarkable and new 
features survive the descent to finite dimension. Mean field is in the 
case of usual, non-disordered systems, giving very good hints about 
the finite dimensional case, but what about systems which offer such 
a series of completely new phenomena? In the last period much 
activity has been devoted to try and clarify this problem. As in many 
murky situations Monte Carlo simulations are playing an important 
role \cite{MC}. Computing corrections to the field theory of the 
problem is also a difficult task, but progresses are being obtained 
\cite{CORREC}. Here in the following we will select the problem of 
ultrametricity, and try to understand how this feature is modified 
when going from mean field to finite dimensional, realistic systems.

The hierarchical solution of mean field spin glasses admits a state structure 
endowed with an ultrametric distance \cite{MPSTV} (for a very good 
discussion of the problem, both introductory and going deep in the 
details of the subject, see \cite{RATOVI}).  Distances among states 
obey the ultrametric inequality we have given before.  Let us consider 
two spin configurations representative of two given states 
\cite{PARJSP}.  One can define the squared distance of two spin 
configurations as

\begin{equation}
  d_{\alpha,\beta}^2 \equiv \frac{1}{4q_{EA}V}
  \sum_{i=1}^V(m_i^\alpha -  m_i^\beta )^2
  =\frac12 \bigl( 1-\frac{q_{\alpha,\beta}}{q_{EA}}  \bigr)\ ,
\end{equation}
where  $q_{\alpha,\beta}\equiv\frac{1}{V}
\sum_{i=1}^V \sigma_i^\alpha   \sigma_i^\beta $ is the overlap of the 
two configurations $C_\alpha$ and $C_\beta$.
Such $d^{2}$ takes the value $0$ when the mutual overlap is exactly $q_{EA}$ 
(the overlap of two configurations in the same state, i.e.  the 
maximum allowed overlap), and the value $1$ when the overlap is $-q_{EA}$.  
This is the distance we will always have in mind in this note.

The main result one obtains in mean field concerns the disorder 
averaged probability distribution for the probability distribution of 
three overlaps.  We consider three equilibrium configurations, $1$, 
$2$ and $3$ of the spin systems (interacting by the same given 
realization of the quenched couplings).
$q_{1,2}$, $q_{2,3}$ and $q_{1,3}$ are the mutual overlaps.
By using the formalism of replica symmetry breaking 
one finds \cite{MPSTV} that

\begin{eqnarray}
&& \overline{
P_{J}(q_{1,2},q_{2,3},q_{1,3})}\nonumber \\
\nonumber &=&
\frac12  P(q_{1,2})x(q_{1,2}) \delta(q_{1,2}-q_{2,3})\delta(q_{1,2}-q_{1,3})\\
\nonumber
&+& \{ \frac12\{P(q_{1,2})P(q_{2,3})\theta(q_{1,2}-q_{2,3})
\delta(q_{2,3}-q_{1,3})\\
&+& \mbox{two permutations} \} \ ,
\end{eqnarray}
where $x(q)\equiv\int_0^qP(q')dq'$ gives the weight of equilateral 
triangles (the other three terms represent triangles with 
the two equal edges longer than the different one).

Further work on the ultrametric features of mean field solution 
\cite{FRPAVI} has clarified the robustness of the ultrametric behavior.
Numerical work on the subject is contained in references 
\cite{UMMC,CIPARI}.

We want to understand what happen in the case of finite dimensional 
systems. It is clear that the problem is a difficult one: finite size 
effects are known to be severe, and the use of a scaling analysis is 
mandatory.

We have used a {\em constrained Monte Carlo} procedure.  For each 
realization of the quenched disordered couplings we have considered 
three configurations of the spin variables, say $C_{\alpha}$, 
$C_\beta$ and $C_\gamma$.  We have fixed the distance of $C_\alpha$ 
from $C_\beta$ and of $C_\beta$ from $C_\gamma$ i.e.  we have kept 
fixed to a constant value the overlaps $q_{\alpha,\beta}\equiv q_{1,2}$ 
and $q_{\beta,\gamma}=q_{2,3}$.  A sensible choice of $q_{1,2}$ and $q_{2,3}$ is 
crucial for the method to give useful results.  The values $q_{1,2}$ and 
$q_{2,3}$ have been kept constant by forbidding spin updates that bring 
the overlap $q_{\alpha,\beta}$ out of the range $q_{1,2}\pm \epsilon$ or 
the overlap $q_{\beta,\gamma}$ out of the range $q_{2,3}\pm \epsilon$.  
For all the simulations discussed in this note we have used 
$\epsilon=0.04$.  A systematic study of the choice of an optimal value 
for $\epsilon$ is contained in \cite{CACCIU}.

By using this procedure we are restricting the phase space: our 
numerical simulations do not investigate the equilibrium properties of 
the full model, but only the sector where in a triple of states two 
distance are fixed. To make the procedure consistent $q_{1,2}$ and $q_{2,3}$ 
have to be chosen in the support of the $P(q)$ of the full model. A 
good choice of the constrained value will make the ultrametric bound 
very different from the triangular bound, making the difference among 
the two phase space structure as clear as possible.

We have studied the $4$ dimensional Edwards-Anderson model, with quenched 
couplings $J=\pm 1$ with probability $\frac{1}{2}$. The Hamiltonian 
$H = \sum_{i,j}\sigma_i J_{ij} \sigma_j$ contains a sum over first 
neighbors. We have chosen the $4d$ (as opposite to $3d$) case because 
here we have a better understanding of the critical behavior 
\cite{CIPARI,MC} ($d=4$ is farer from the lower critical 
dimension, and the critical behavior is clearer).  We have been 
working at $T=1.4$, i.e.  $T\simeq 0.7 T_c$, where we are already in 
the broken phase and the $P(q)$ has a clear non-trivial structure, but 
we are still able to completely thermalize the non-constrained system, 
at least on small lattices \cite{CACCIU}.

We have used lattices of volume $V=L^4$ with $L=3$, $4$, $5$, $6$, $7$ 
and $8$.  $L=8$ was the maximum lattice size we felt sure we were able 
to thermalize.  We have averaged the $P_J(q)$ over different 
realizations of the quenched disordered couplings $J$; in both the 
numerical experiments we will describe in the following the number of 
samples $N_J$ has been a maximum of $2000$ for the smallest lattice 
($L=3$) and a minimum of $100$ for the largest lattice, $L=8$ (with 
intermediate $N_J$ values for the intermediate lattice sizes).

We will give more details about the thermalization of the samples, 
that is for this kind of numerical experiment a crucial issue. 
We have been very careful about this point, and all the 
data we present here have passed detailed thermalization tests 
\cite{CACCIU}. For all lattice size we have used an annealing 
schedule; starting from a random configuration we have cooled the 
systems from $T=2.4$ down to $T=1.4$ at steps $\Delta T = 0.1$ (for 
the smaller $L$ values) or $0.2$ (for the larger ones). At each $T$ 
value we have ran from $25000$ sweeps ($L=3$) to $80000$ sweeps ($L=8$). 
Once at $T=1.4$, after the usual thermalization steps, we have ran 
from half a million ($L=3$) to eight hundred thousand sweeps ($L=8$) 
for measuring.

Our main results have been obtained by fixing

\begin{equation}
  q_{\alpha,\beta}\equiv q_{1,2} = q_{\beta,\gamma} 
  = \frac{2}{5} q_{EA}
  \equiv \overline{q}\ ,
\end{equation}
where by $q_{EA}$ we mean the infinite volume values as estimated for 
example in \cite{CIPARI} ($.54$). That means we are fixing 
$\overline{q}=.21$ for all the volume values we investigate.
In this condition the triangular inequality obliges
the measured $q$, the third side of the triangle, to obey

\begin{equation}
  q \ge - \frac75  q_{EA}\ ,
\end{equation}
while an ultrametric distance would imply

\begin{equation}
  q \ge  \frac25  q_{EA}\ .
\end{equation}
It is clear that the two bounds are very different.  Obviously in 
both cases in the infinite volume limit the measured $q$ will be 
smaller than $q_{EA}$.

\begin{figure}
\epsfig{file=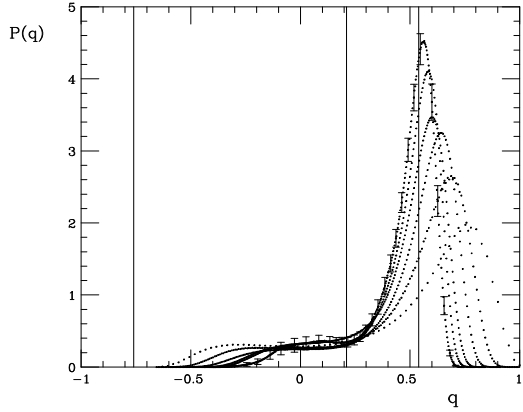,width=300pt,angle=90}
\protect\caption{
$P(q)$ of the free overlap measured in the course of our constrained 
Monte Carlo  as a function of $q$, for $L$ values going from $3$ 
(lowest curve, ending on the left closest to $q=0$) to $L=8$
(highest curve, ending on the left farest from $q=0$). }
\protect\label{FFIG1}
\end{figure}

In fig.  (\ref{FFIG1}) the vertical line on the left, at $q\simeq 
-0.75$, depicts the bound given from the triangular inequality.  The 
second vertical line, at $q\simeq .21$, depicts the ultrametric bound, 
while the vertical line on the right, at $q_{EA}$, is the upper bound 
for infinite volume.  The probability distributions of the 
measured $q$ value (the overlap among configuration $C_\alpha$ and 
configuration $C_\beta$, see before) for the $6$ $L$ values, from 
$L=3$ to $L=8$.  The $L=3$ $P(q)$ is the one with the smaller peak, 
farest on the right, that end last on the left: $P(q)$ for increasing 
$L$ values have higher peaks, and end more and more at $q$ values 
close to zero.  It is clear that already on small lattices the 
distribution is far from the triangular bound.

The probability for a measured 
distance $q$ not to be ultrametric (i.e. one minus
the normalized area $S^U$ of 
the $P(q)$ integrated inside the ultrametric bound)
decreases fast with the lattice size. On a $L=8$ lattice half of the 
configurations are ultrametric (and indeed the big violation is from 
configurations with $q>q_{EA}$, which we expect from normal Monte Carlo 
runs to disappear in the continuum limit).

\begin{figure}
\epsfig{file=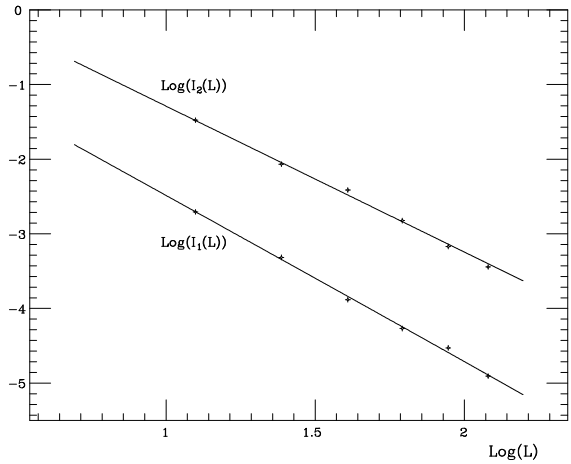,width=300pt,angle=90}
\protect\caption{
The integral $I^{L}$ as a function of $L$, in double log scale.
The lower points
are for the case where we have fixed $q_{1,2}= q_{2,3}$, the upper points
where $q_{1,2}\ne q_{2,3}$ (see the text).}
\protect\label{FFIG2}
\end{figure}

In fig. (\ref{FFIG2}) we plot the value of the integral

\begin{eqnarray}
  \nonumber
  &I^L& \equiv \int_{-1}^{q_{min}}(q(L)-q_{min})^2 P(q)dq\\
   &+&
  \int_{q_{MAX}}^{1}(q(L)-q_{MAX})^2 P(q)dq \ ,
\end{eqnarray}
in log-log scale. Here $q_{min}=q_{1,2}$ and $q_{MAX}=q_{EA}$.
The lower points ($I_1$) are for the case we are discussing here, 
the upper ones for the case where we fix $q_{1,2}\ne q_{2,3}$ (see later). 
The straight line is our very good best fit to a power behavior, that 
gives

\begin{equation} 
  I^{L} \simeq (-.0001\pm .0005) + (0.76\pm 0.03)L^{-2.21\pm 0.04}\ . 
\end{equation} 
The integral goes to zero in the infinite volume limit, as we 
would expect for an ultrametric structure.  It is remarkable that the 
asymptotic value is estimated to be so close to zero, and that the 
estimated exponent is very close to the $\frac83$ one expects from the 
results of \cite{FRPAVI}.

A few comments are in order.  There are two different and important 
effects in fig.  (\ref{FFIG1}).  On the one side the number of 
configurations with large $q$ ($q>q_{EA}^{V=\infty}$) decreases with 
increasing $L$.  These are the kind of finite size effects that 
normally one studies.  Such finite size effects are already known to 
be quite large in quenched disordered systems: even for the SK model 
it is quite difficult to get a good numerical determination of the 
position of the peak of $P(q)$ in the infinite volume limit.  We 
confirm that in our simulation.  On the other side ($q<0$) things are 
different.  Already on a very small lattice very few configurations 
are allowed in the region that is allowed by the triangular inequality 
but ultrametrically forbidden.  This region is systematically reduced 
when increasing the lattice size (it is basically halved when going 
from $L=3$ to $L=8$).

We have also fitted the position of the peak of $P(q)$, $q^{(\infty)}_{MAX}$
with an $L$-dependent power law. Our best fit is very good and gives

\begin{equation}
  q^{(L)}_{MAX} = (0.31\pm0.09) + (0.85\pm 0.03) L^{-0.59\pm 0.15}\ ,
\end{equation}
with a value of $q^{(\infty)}_{MAX}$ just in the center of the allowed 
ultrametric region.

These results look very positive. Already on the lattice sizes one can 
equilibrate with a numerical simulation (in a sector of 
the phase space) that took a few months of 
workstation time one clearly sees the increasing domination of 
ultrametric sample couples. Obviously we do not know if in the 
infinite volume limit there will be a completely ultrametric 
structure (we do not have a priori reasons to be sure of that) or if 
ultrametricity will be realized on finite dimensional spin glasses 
only as  a dominance of states close to ultrametric behavior: 
certainly, we are showing here that the ultrametric sector of the 
phase space is very important.

Maybe the most important problem we are detecting is the one of large 
finite size effects.  That was already known, and we confirm it here: 
for example $q^{(L)}_{EA}$ converges only very slowly to 
$q^{(\infty)}_{MAX}$.

\begin{figure}
\epsfig{file=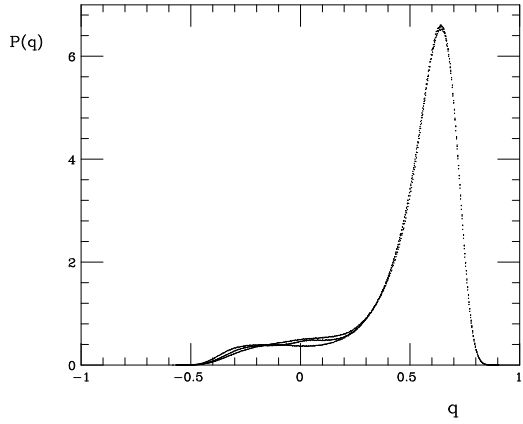,width=300pt,angle=90}
\caption{
The three curves (basically coinciding in the plot) are for $P(q)$ in the first, 
the second and the last  third of the run (after the annealing scheme 
described in the text), $L=5$, $100$ samples.}
\protect\label{FFIG3}
\end{figure}

Thermalization is a key problem.  In a constrained dynamics like the 
one we are using is very difficult to ascertain thermalization.  The 
usual criterion of checking that the dynamical $P_d(q)$ (i.e.  the one 
where the overlap is defined from $\lim_{t_1\to 
\infty}\sigma(t)\sigma(t+t_1)$) should coincide with the equal time 
$P_e(q)$ (where the overlap is from the evolution of two different 
systems, i.e.  from $\sigma(t)\tau(t)$) is not useful here (since in 
our constrained Monte Carlo method we do not have an equivalent 
dynamical quantity $P_d(q)$ to compare with).  Also the symmetry 
$\sigma\to-\sigma$ is not a good symmetry here, and the symmetry of 
$P(q)$ cannot be used to check thermalization.  Because of that we 
have tried to stay on the very safe side.  In figure (\ref{FFIG3}) we 
plot $P(q)$ for $L=5$ ($100$ samples) for the first, the second and 
the third third of the MC sweeps (after the annealing schedule).  The 
three curves are basically identical, making as confident that we are 
having no thermalization problems.

\begin{figure}
\epsfig{file=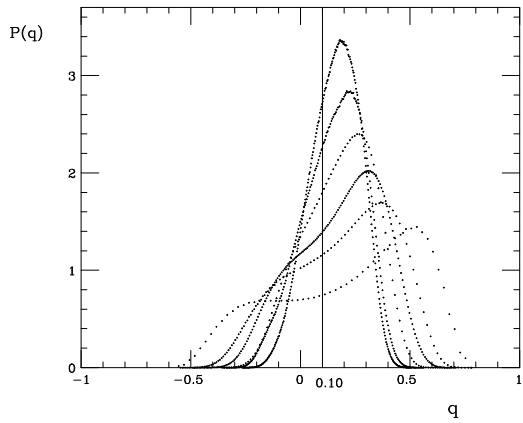,width=300pt,angle=90}
\protect\caption{
As in figure (\protect\ref{FFIG1}), but for $q_{1,2}=\frac45 q_{EA}$ and 
$q_{2,3}=\frac15 q_{EA}$.}
\protect\label{FFIG4}
\end{figure}

We have also analyzed a different situation, in which we have set $q_{1,2} 
\ne q_{2,3}$.  In this case we have used runs of length similar to the one 
of the previous case, a similar number of samples, same $L$ values and 
$T=1.4$.  We have fixed $q_{1,2} = \frac45 q_{EA}$ and $q_{2,3} = \frac15 
q_{EA}$.  In this case an ultrametric behavior implies that in the 
infinite volume limit $P(q)=\delta(q-\frac15 q_{EA})$, centered in 
$\frac15 q_{EA}\simeq 0.10$.  We report in fig.  (\ref{FFIG4}) the 
$P(q)$ for this situation.  A power fit for the position of the peak 
in the infinite volume limit gives as a preferred value 
$q^{(\infty)}_{MAX}=(0.10\pm 0.03)$, right on the ultrametric point.  
Also in this case we are getting a strong indication toward the 
presence of ultrametric features in the state space of finite 
dimensional spin glasses.

In fig. (\ref{FFIG2}) we also plot the value of the integral $I^L$ for 
this second case ($I_2$).
Here $q_{min}=q_{MAX}$ is the location where we expect a delta 
function to be built in the infinite volume limit, and the integral 
just goes from $-1$ to $1$.
The straight line is again the best fit to a power behavior, that 
gives

\begin{equation} 
  I^{L} \simeq (-.000\pm .002) + (1.95\pm 0.08)L^{-1.95\pm 0.04}\ . 
\end{equation} 
Also here the integral seems to be going to zero 
in the infinite volume limit.

We are grateful to Felix Ritort for sharing with us some ideas about 
this problem.
We also thank Peter Young for an interesting conversation.

\end{document}